\documentclass[pra,twocolumn,floatfix,superscriptaddress]{revtex4-1}
\usepackage{dcolumn,amsmath}
\usepackage{graphicx}
\usepackage{bm}
\usepackage{hyperref}

\usepackage{color}
\definecolor{BLUE}{rgb}{0.0,0.0,1.0}

\usepackage{soul}
\soulregister\cite7
\soulregister\ref7

\usepackage{dcolumn}
\usepackage{siunitx}
\usepackage{multirow}
\usepackage{bigdelim}

\usepackage{float}

\usepackage{hyperref}
\hypersetup{colorlinks=true, citecolor=blue, urlcolor=blue, linkcolor=blue}

\usepackage[dvipsnames]{xcolor}

\usepackage{accents}

\usepackage{xcolor}

\begin{document}

\title{Isotope-shift factors with quantum electrodynamics effects for many-electron systems: A study of the nuclear charge radius of $^{26m}$Al}

\date{April 20, 2024}

\begin{abstract}
A method for calculating the field shift contribution to isotope shifts in many-electron atoms, incorporating quantum electrodynamics (QED) effects, is introduced. We also implement the model QED approach to incorporate QED contribution to the nuclear recoil effect at the high-order correlation effects treatment level. The proposed computational scheme is used to revise the value of the root-mean-square (rms) nuclear charge radius of the isomer of aluminium-26, $^{26m}$Al. This radius is important for the global analysis of the $V_{ud}$ element of the Cabibbo-Kobayashi-Maskawa matrix. The difference in mean-square nuclear charge radii of $^{27}$Al and $^{26m}$Al, obtained  by combining the calculated atomic factors with recently measured isotope shift (IS) of the $3s^23p~^2P_{3/2} \to 3s^24s~^2S_{1/2}$ transition in Al, is $0.443(44)(19)~{\rm fm}^2$, where the first and second uncertainties are experimental and theoretical ones, respectively. The latter is reduced by a factor of 4 with respect to the previous study. Using this value and the known value of the rms charge radius of $^{27}$Al, the resultant value $R_c(^{26m}$Al) = 3.132(10)~fm is obtained. With the improved accuracy of the calculated IS factors the error in $R_c(^{26m}$Al) is now dominated by the experimental uncertainty. 
Similar revision of rms charge radii is made for the $^{28}$Al, $^{29}$Al, $^{30}$Al, $^{31}$Al and $^{32}$Al isotopes using existing IS measurements. Additionally, atomic factors are computed for 
the {$3s^23p~^2P_{3/2} \to 3s^24s~^2S_{1/2}$}, {$3s^23p~^2P_{1/2} \to 3s^25s~^2S_{1/2}$}  and {$3s^23p~^2P_{3/2} \to 3s^25s~^2S_{1/2}$} transitions in Al, which can be used in future experimental studies.
\end{abstract}

\author{Leonid V.\ Skripnikov}
\email{skripnikov\_lv@pnpi.nrcki.ru,\\ leonidos239@gmail.com}
\homepage{http://www.qchem.pnpi.spb.ru}
\affiliation{Petersburg Nuclear Physics Institute named by B.P. Konstantinov of National Research Centre
``Kurchatov Institute'', Gatchina, Leningrad District 188300, Russia}
\affiliation{Saint Petersburg State University, 7/9 Universitetskaya nab., St. Petersburg, 199034 Russia}

\author{Sergey D.\ Prosnyak}
\affiliation{Petersburg Nuclear Physics Institute named by B.P. Konstantinov of National Research Centre
``Kurchatov Institute'', Gatchina, Leningrad District 188300, Russia}
\affiliation{Saint Petersburg State University, 7/9 Universitetskaya nab., St. Petersburg, 199034 Russia}

\author{Aleksei V. Malyshev}
\affiliation{Saint Petersburg State University, 7/9 Universitetskaya nab., St. Petersburg, 199034 Russia}
\affiliation{Petersburg Nuclear Physics Institute named by B.P. Konstantinov of National Research Centre
``Kurchatov Institute'', Gatchina, Leningrad District 188300, Russia}

\author{Michail Athanasakis-Kaklamanakis}
\affiliation{Centre for Cold Matter, Imperial College London, SW7 2AZ London, United Kingdom}
\affiliation{KU Leuven, Instituut voor Kern- en Stralingsfysica, B-3001 Leuven, Belgium}

\author{Alex Jose Brinson}
\affiliation{Department of Physics, Massachusetts Institute of Technology, Cambridge, MA 02139, USA}

\author{Kei Minamisono}
\affiliation{Facility for Rare Isotope Beams, Michigan State University, East Lansing 48824, USA}
\affiliation{Department of Physics and Astronomy, Michigan State University, East Lansing 48824, USA}

\author{Fabian C. Pastrana Cruz}
\affiliation{Department of Physics, Massachusetts Institute of Technology, Cambridge, MA 02139, USA}

\author{Jordan Ray Reilly}
\affiliation{Department of Physics and Astronomy, The University of Manchester, Manchester M13 9PL, United Kingdom}

\author{Brooke J. Rickey}
\affiliation{Facility for Rare Isotope Beams, Michigan State University, East Lansing 48824, USA}
\affiliation{Department of Physics and Astronomy, Michigan State University, East Lansing 48824, USA}

\author{Ronald. F. Garcia Ruiz}
\affiliation{Department of Physics, Massachusetts Institute of Technology, Cambridge, MA 02139, USA}
\affiliation{Laboratory for Nuclear Science, Massachusetts Institute of Technology, Cambridge, MA 02139, USA}


\maketitle

\section{Introduction}

Accurate determination of nuclear charge radii serves as sensitive examination of various elements of nuclear structure~\cite{Yang2022_exotic}, providing an important benchmark for the development of  nuclear models~\cite{Garcia2016,deGroote2020,Koszorus2021,Goodacre2021}. Precise charge radii measurements can also be used to constrain the parameters of the equation of state of nuclear matter~\cite{Pineda2021,kon24}.

The nuclear charge radii of some isotopes can be used as an important component to test fundamental particle models. The Cabibbo-Kobayashi-Maskawa (CKM) matrix plays a central role in describing the quark-flavour mixing via the weak interaction in the Standard Model (SM). According to the SM, the CKM matrix has to be unitary. However, this should be verified experimentally. Non-unitarity can be a manifestation of new physics beyond the SM. Significant efforts are undertaken to verify the property of unitarity~\cite{Hardy:2020}. The deviation from the top-row unitarity of the CKM matrix can be characterized by the $\Delta_{\rm CKM}=1-(|V_{ud}|^2+|V_{us}|^2+|V_{ub}|^2)$ parameter, which is expected to be zero in the unitary case. The value of $V_{ud}$ can be derived from the global analysis of superallowed $0^+ \to 0^+$ nuclear $\beta$ decay of certain isotopes~\cite{Hardy:2020}. Among them, the superallowed $\beta$ decay of $^{26m}$Al isomer is of key importance, as it has the smallest nuclear structure-dependent corrections~\cite{Hardy:2020}. Several radiative corrections have to be calculated to connect the experimentally measured $ft$ value, characterizing the superallowed $\beta$ decay, with the vector coupling constant $G_V$ and the $V_{ud}$ matrix element. One of these corrections is the isospin-symmetry-breaking constant, which depends on the nuclear mean-square (ms) charge radius.

Nuclear ms charge radii can be determined through measurements of the spectral line isotope shifts (IS) between different isotopes of a given atom~\cite{Yang2022_exotic}. The corresponding technique also becomes a testing ground to probe higher-order nuclear structure effects and new long- and intermediate-range interactions~\cite{New1,New2,New3,New4}. IS arises mainly due to differences in the masses and nuclear electrostatic potentials of the isotopes, i.e., due to nuclear recoil and field shift effects~\cite{Yang2022_exotic}. The purpose of the atomic electronic structure theory is to calculate these effects and connect experimental observables with fundamental nuclear properties. A very accurate description of IS effects can be achieved in highly charged ions, where electron correlation effects are suppressed and methods of rigorous quantum electrodynamics (QED) are successfully applied (see, e.g.,  Refs.~\cite{Zubova:2016,Adkins:2007:042508,Blundell:1993,Anisimova:2022} and references therein). For neutral many-electron atoms, the electronic structure problem becomes very challenging as electron correlation effects are not suppressed and significantly contribute to the atomic factors that describe the IS effects. As a result, approximate treatment of electron correlations usually provides the dominant source of uncertainty in the extracted ms charge radii, negating the experimental achievements. Electron correlation effects also make it difficult to rigorously calculate QED corrections to IS atomic factors, and attempts to treat them in neutral many-electron systems usually result in order-of-magnitude estimates at best.

Until recently~\cite{Plattner:2023}, the value of the ms charge radius for $^{26m}$Al was known only from extrapolations~\cite{Towner:2002,Towner:2008}. The IS study~\cite{Plattner:2023} resulted in a significant reexamination of the nuclear ms charge radius of $^{26m}$Al. However, the extraction of 
the radius 
from these measurements requires input from atomic theory ~\cite{Heylen:2021, Plattner:2023}. The 
IS
between $^{26m}$Al and $^{27}$Al was measured using collinear laser spectroscopy of the $3s^23p~^2P_{3/2} \to 3s^24s~^2S_{1/2}$ transition \cite{Plattner:2023}, and the uncertainty in the extracted ms charge radii difference between $^{26m}$Al and $^{27}$Al was strongly dominated by the atomic theory calculations.

In the present paper, we undertake the accurate determination of atomic factors, which allows for a significant reduction in the theoretical and, hence, total uncertainty of the ms charge radius of $^{26m}$Al as well as $^{28}$Al, $^{29}$Al, $^{30}$Al, $^{31}$Al and $^{32}$Al isotopes. To achieve this, we 
develop a computational scheme to
perform high-precision relativistic electronic structure calculations within the coupled cluster theory with inclusion of up to quadruple excitation amplitudes. Furthermore, we develop an effective approach to incorporate leading-order 
QED
effects for these problems with precision comparable to that previously available only for similar transitions in highly charged ions, treated within the rigorous QED approaches.

\section{\label{sec:theory}Theory}

\subsection{Leading order isotope shift effect}

The isotope shift (IS) of the atomic transition energy $\Delta \nu^{M', M} = \nu^{M'}-\nu^{M}$ can be parameterized as follows:
\begin{equation}
\label{freq}
    \Delta \nu^{M', M} = (k_{\rm NMS}+k_{\rm SMS})\left(\frac{1}{M'}-\frac{1}{M}\right)+F\delta\left\langle r^2 \right\rangle^{M', M}.
\end{equation}
Here $k_{\rm NMS}$ and $k_{\rm SMS}$ are the normal and specific mass shift constants, characterising the nuclear recoil effect, $M$ and $M'$ are the masses of the considered pair of isotopes. Constant $F$ characterizes the field shift effect due to change of the nuclear ms charge radius $\delta\left\langle r^2 \right\rangle^{M', M}=\left\langle r^2 \right\rangle(M')-\left\langle r^2 \right\rangle(M)$. 
In the present work the field shift (FS) constant $F$ is defined as 
\begin{equation}
\label{FSdef}
F= \frac{d\, \nu}{d \left\langle r^2 \right\rangle} , 
\end{equation}
where $\nu$ is the electronic transition energy, and the derivative is calculated at the point $r_{\rm rms}=\sqrt{\langle r^2 \rangle}=3.061$~fm, which corresponds to $^{27}$Al~\cite{ANGELI201369,Heylen:2021}.

Provided the isotope shift is known, the difference $\delta\left\langle r^2 \right\rangle^{M', M}$ can be extracted if one knows the values of the atomic factors $k_{\rm NMS}$, $k_{\rm SMS}$, and $F$. These values can be obtained theoretically through the solution of the electronic many-body problem. To incorporate relativistic effects, one can use the Dirac-Coulomb(-Breit) Hamiltonian,
\begin{equation}
\label{EQ:HAMILTONIAN}
 H = \Lambda_{+} \left( \sum\limits_i\, \left[ c\, \boldsymbol{\alpha}_i \cdot {\mathbf{p}}_i + \beta_i c^2 
+H_{\rm nuc}(i) \right]
+ V_{ee} \right) \Lambda_{+},
\end{equation}
where $\alpha$, $\beta$ are the Dirac matrices, $\mathbf{p}_i$ denotes the momentum of the $i$-th electron, $H_{\rm nuc}$ is the electron-nucleus interaction operator, summation is over all electrons, $\Lambda_{+}$ is the projector on the positive-energy states (obtained in the Dirac-Hartree-Fock (DHF) procedure), and $V_{ee}$ is the electron-electron interaction. In the present work we considered the Coulomb and Coulomb-Breit interelectronic interaction operators.

Within the Breit approximation and to the lowest order in the electron-to-nucleus mass ratio $m/M$ the mass shift effect can be calculated using the following relativistic operators~\cite{shabaev1985mass, palmer1987reformulation, shabaev1988nucl, shabaev1994relativistic}:
\begin{equation}
\label{opNMS}
    H_{\rm NMS} = \frac{1}{2M}\sum_i\left(\mathbf{p}_i^2-\frac{\alpha Z}{r_i}\left[ \bm{\alpha}_i+\frac{(\bm{\alpha}_i\cdot \mathbf{r}_i)\mathbf{r}_i}{r_i^2}\right]\cdot\mathbf{p}_i\right),
 \end{equation}
\begin{equation}
\label{opSMS}
    H_{\rm SMS} = \frac{1}{2M}\sum_{i \neq k}\left(\mathbf{p}_i\cdot\mathbf{p}_k-\frac{\alpha Z}{r_i}\left[ \bm{\alpha}_i+\frac{(\bm{\alpha}_i\cdot \mathbf{r}_i)\mathbf{r}_i}{r_i^2}\right]\cdot\mathbf{p}_k\right),
\end{equation}
where $Z$ is the proton number and $\mathbf{r}_i$ is the coordinate of the $i$-th electron. Note that $H_{\rm SMS}$ is a two-electron operator. The constants $k_{\rm NMS}$ and $k_{\rm SMS}$ (up to a multiplier of $1/M$) can be calculated as expectation values of the operators (\ref{opNMS}) and (\ref{opSMS}) with the wave functions obtained using the Dirac-Coulomb-Breit Hamiltonian (\ref{EQ:HAMILTONIAN}).

\subsection{\label{SecFSqed} QED contribution to the field shift}

In most electronic structure studies of many-electron atoms (with more than ten electrons), the Dirac-Coulomb-Breit approximation appears to be sufficient,
as the uncertainty due to the treatment of electron correlation effects is usually quite significant. However, for the precise study of highly charged ions (HCI), where electron-electron correlations are strongly suppressed in comparison to the electron-nucleus interaction, the treatment of quantum electrodynamics effects becomes crucial, see Ref. \cite{Sapirstein:2008:25, Glazov:2011:71, Shabaev:2018:60, Indelicato:2019:232001, Shabaev:2024:94:inbook,Blundell:1993} and references therein. The rigorous consideration of the QED contribution to the field shift constant was recently undertaken for HCIs in Refs. \cite{Zubova:2016, King:2022:43}. The application of such methods to neutral many-electron systems, where electron correlation effects are not suppressed as in HCIs and have to be considered to high orders, is challenging. For these reasons, a prospective approach is the use of effective (model) QED operators.

Let us suppose that QED contribution is modeled by some effective operator $H^{\rm QED}(\mathbf{r}, \langle r^2 \rangle)$, which depends on the electron coordinate $\mathbf{r}$ and the ms nuclear charge radius $\langle r^2 \rangle$. In the first order, the QED contribution to the total electronic energy is given by the expectation value
\begin{equation}
E^{\rm QED}(\langle r^2 \rangle) = \langle \Psi^{\langle r^2 \rangle} | \sum_i H^{\rm QED}(\mathbf{r}_i, \langle r^2 \rangle) | \Psi^{\langle r^2 \rangle}\rangle,   
\label{EQED}
\end{equation}
where $\Psi^{\langle r^2 \rangle}$ is the many-electron wave function calculated using Hamiltonian (\ref{EQ:HAMILTONIAN}), which depends on the ms nuclear charge radius through the $H_{\rm nuc}$ operator. Therefore, the contribution of the QED effects to the FS constant, $F^{\rm QED}$, can be calculated as follows:
\begin{equation}
\label{FQED}
F^{\rm QED} = \frac{d \langle \Psi^{\langle r^2 \rangle} | \sum_i H^{\rm QED}(\mathbf{r}_i, \langle r^2 \rangle) | \Psi^{\langle r^2 \rangle} \rangle}{d \langle r^2 \rangle}.
\end{equation}

The leading QED contributions are given by the one-loop vacuum polarization (VP) and self-energy (SE):
\begin{equation}
    H^{\rm QED} = H^{\rm VP}+H^{\rm SE}.
    \label{eq:h_se_vp}
\end{equation}
The dominant part of the VP interaction is represented by the Uehling potential, for which one can use the analytical expression for a finite nucleus case:
\begin{widetext}
\begin{equation}
\label{VPoper}
H^{\rm Ueh}(r)=-\frac{2}{3}\frac{\alpha^2}{r}\int_0^{\infty}dr'\int_1^{\infty}dt\sqrt{t^2-1}\Big(
\frac{1}{t^3}+\frac{1}{2t^5}\Big) \rho_{\rm nuc} (r') r' \Big(
e^{-2t|r-r'|/\alpha}-e^{-2t(r+r')/\alpha}
\Big) \ ,
\end{equation}
\end{widetext}
where $\rho_{\rm nuc}$ is the nucleus charge density normalized according to $\int \rho_{\rm nuc} (r)\, d^3{\mathbf r}=Z$, and $\langle r^2 \rangle=\int r^2 \rho_{\rm nuc} (r)\, d^3{\mathbf r}/\int \rho_{\rm nuc} (r)\, d^3{\mathbf r}$. A simple expression can be obtained for the homogeneous charge distribution model~\cite{Ginges:2016}. Due to the known explicit expression, one can directly calculate the VP contribution to $F^{\rm QED}$ within Eq.~(\ref{FQED}). The contribution of the higher-order part of the VP interaction, given by the Wichmann-Kroll potential, is negligible. The discussion of this term is beyond the scope of the present work.

The treatment of the SE QED effects is more challenging. Several approximate expressions have been suggested to take into account this effect on binding energies of HCIs and atoms~\cite{Shabaev:13,Flambaum:2005,Pyykko:2003,Thierfelder:2010,Schwerdtfeger:2017,LeimbachAt:2020,Indelicato:1990,Tupitsyn:2013,Lowe:2013,Sunaga:2022}. The main idea is the scaling of the SE contribution of interest to the Lamb shift result for the Coulomb potential. In Ref.~\cite{Shabaev:13}, in contrast to many other suggestions, an expression for $H^{\rm SE}$ was proposed, that is based on the use of diagonal and off-diagonal matrix elements of the self-energy operator for H-like ions. The scaling is possible due to the fact that the dominant part of the QED effects comes from the vicinity of the nucleus and the proportionality property of the atom radial wave functions having different principal quantum numbers $n$ and the same relativistic quantum number $\kappa=(-1)^{j+l+1/2}(j+1/2)$.

In Ref.~\cite{Skripnikov:2021a}, a slightly different form of the SE effective operator from Ref.~\cite{Shabaev:13} was introduced and implemented, which can be conveniently used in molecular and atomic studies. The approach uses the same SE matrix elements calculated in Ref.~\cite{Shabaev:13} for H-like systems. The model SE operator from Ref.~\cite{Skripnikov:2021a} can be written in the following form:
\begin{equation}
    H^{\rm SE} =\sum_{k,k',ljm} |h_{kljm}\rangle X_{kljm,k'ljm} \langle h_{k'ljm}|,
\label{Xmod}    
\end{equation}
where $h_{kljm}(\mathbf{r})$ are an orthonormalized set of numerically linearly independent functions, being linear combinations of functions of the type:
\begin{equation}
  \widetilde{h}_{nljm}(\mathbf{r})=\eta_{nljm}(\mathbf{r}) \theta(R_{\rm cut}-|\mathbf{r}|).
\label{hfuns}  
\end{equation}
Here  $\theta(R_{\rm cut}-|\mathbf{r}|)$ is the Heaviside step function, $\eta_{nljm}(\mathbf{r})$ are H-like functions (with principal quantum numbers $n \le 5$), and $R_{\rm cut}$ is a small radius, which can be varied slightly to study the stability of the results obtained. $X_{kljm,k'ljm}$ in Eq.~(\ref{Xmod}) are matrix elements of the SE operator over the $h_{kljm}$ functions and, for the reasons mentioned above, to good accuracy they are the linear combinations of the corresponding matrix elements over the H-like functions. Note, that the SE operator is diagonal in $ljm$. The expressions for both diagonal and off-diagonal matrix elements of SE operator can be derived, e.g., within the two-time Green's function method \cite{Shabaev:02a}:
\begin{equation}
\label{SEme}
    \langle \eta_{n} | \frac{1}{2} (\Sigma^{\rm SE}(\epsilon_{n}) +  \Sigma^{\rm SE}(\epsilon_{n'})) | \eta_{n'} \rangle,
\end{equation}
where $\Sigma^{\rm SE}(\epsilon_{n})$ is the renormalized SE operator. The model SE operator introduced in Ref.~\cite{Shabaev:13} or its variant proposed in Ref.~\cite{Skripnikov:2021a} can be used to calculate the SE contribution to the total electronic energy in various systems~\cite{Shabaev:13, Tupitsyn:2016:253001, Schwerdtfeger:2017, Yerokhin:2017:042505:2017:069901:join_pr, Machado:2018:032517, Si:2018:012504, Muller:2018:033416, Kaygorodov:2019:032505, Zaytsev:2019:052504, Shabaev:2020:052502, Skripnikov:2021a, Kaygorodov:2021, Savelyev:2022:012806, Kaygorodov:2022:062805, Savelyev:2023:042803,Skripnikov:2021b,AthanasakisRaFPinning:2023, Skripnikov:2024a}. However, they cannot be directly applied to evaluate the SE contribution to the FS constant (\ref{FQED}).

Let us now consider how one can calculate this contribution for many-electron systems using existing well-developed methods to treat electron correlation effects. According to Eqs.~(\ref{EQED}), (\ref{eq:h_se_vp}), and (\ref{Xmod}), the SE contribution to the total electronic energy can be expressed as follows:
\begin{align}
E^{\rm SE}(\langle r^2 \rangle) = \langle \Psi^{\langle r^2 \rangle} | \sum_i H^{\rm SE}(\mathbf{r}_i, \langle r^2 \rangle) | \Psi^{\langle r^2 \rangle}\rangle \\
=\sum_{p,q} 
X_{p,q}^{\langle r^2 \rangle}
D_{p,q}^{\langle r^2 \rangle},
\label{ESE}    
\end{align}
where, for brevity, we replaced the set of indices $kljm$ by one index $p$ (or $q$) for elements of the matrix $||X||$ in Eq.~(\ref{Xmod}). We also explicitly indicated the dependence of these matrix elements on the ms nuclear charge radius, and $D_{p,q}^{\langle r^2 \rangle}$ are elements of the one-particle density matrix in the same basis of functions $h_{p}$, which corresponds to the electronic wave function $\Psi^{\langle r^2 \rangle}$.
Now we have:
\begin{align}
\label{Fse}
    F^{\rm SE} = \sum_{p,q} 
\frac{d X_{p,q}^{\langle r^2 \rangle}}
    {d \langle r^2 \rangle} D_{p,q}^{\langle r_0^2 \rangle} 
    +\sum_{p,q} 
    X_{p,q}^{\langle r_0^2 \rangle}
 \frac{d D_{p,q}^{\langle r^2 \rangle}}{d \langle r^2 \rangle},    
\end{align}
where $\langle r_0^2 \rangle$ is the ms nuclear charge radius corresponding to reference isotope and all derivatives are taken at this point.
The derivative of the density matrix pertains to the electronic problem and can be solved using techniques such as numerical differentiation. Therefore, the second term on the right-hand side of Eq.~(\ref{Fse}) can be readily calculated using existing techniques.

To calculate SE contribution due to the first term of Eq.~(\ref{Fse}), we have to work out a procedure for computing the derivative of the SE matrix elements with respect to the nuclear ms charge radius. In Refs.~\cite{Milstein:2003,Milstein:2004,Yerokhin:2011b,Ulrich:2003,Pachucki:1993,Milstein:2002,Eides:1997} the nuclear size dependence of the diagonal matrix elements of the  SE operator over the  functions of H-like ions has been explored and parameterized as follows for $ns, np_{1/2}$ and $np_{3/2}$ states:
\begin{eqnarray}
\label{SEnscorr}
    \Delta E^{\rm SE}= \langle \eta_{nlj} | \Sigma^{\rm SE} | \eta_{nlj}  \rangle \nonumber \\
    = \frac{\alpha}{\pi} \Delta E_{\rm NS}(nl\frac{1}{2},R) G_{\rm NSE}(nlj,R),
\end{eqnarray}
where we have omitted the total angular momentum projection in the index of the function $\eta$, $R=\sqrt{\frac{5}{3}}\sqrt{\langle r^2 \rangle}$, $\Delta E_{\rm NS}$ is the nuclear size (NS) correction to the Dirac energy for the considered state $\eta_{nlj}$, and $G_{\rm NSE}$ is a slowly varying function of $Z$ and $R$~\cite{Yerokhin:2011b}.
Note that we use the $nlj$ notation instead of the $njl$ one employed in Ref.~\cite{Yerokhin:2011b}.
Let us consider the derivative of the matrix element~(\ref{SEnscorr}):
\begin{widetext}
\begin{equation}
    \label{SEderiv}
   \frac{d \Delta E^{\rm SE}(nlj,R)}{d \langle r^2 \rangle} \bigg|_{R_0} 
   =\Delta E^{\rm SE}(nlj,R_0) 
   \left[
   \frac{\frac{d \Delta E_{\rm NS}(nl\frac{1}{2},R)}{d \langle r^2 \rangle}\bigg|_{R_0}}{\Delta E_{\rm NS}(nl\frac{1}{2},R_0)} +\frac{\frac{d G_{\rm NSE}(nlj,R)}{d \langle r^2 \rangle}\bigg|_{R_0}}{G_{\rm NSE}(nlj,R_0)}
   \right],   
\end{equation}
\end{widetext}
where $R_0=\sqrt{\frac{5}{3}}\sqrt{\langle r_0^2\rangle}$.
Eq.~(\ref{SEderiv}) is of crucial practical importance. $\Delta E^{\rm SE}(R_0)$ is the difference of the SE matrix elements calculated for the finite and point nuclei, that is the NS correction to the SE contribution. The NS corrections for both diagonal and off-diagonal SE matrix elements, evaluated for H-like functions, were calculated and tabulated in Ref.~\cite{Shabaev:13}. By considering the formulas compiled in Ref.~\cite{Yerokhin:2011b}, one can realize that the expression in square brackets in Eq.~(\ref{SEderiv}) is almost independent on the principal quantum number $n$ in the leading order (see below). Using the analytical expressions for $\Delta E_{\rm NS}(ns,R)$ and $\Delta E_{\rm NS}(np_{1/2},R)$~\cite{ShabaevNS:1993}, one obtains:
\begin{equation}
\label{Ederiv}
\frac{\frac{d \Delta E_{\rm NS}(ns,R)}{d \langle r^2 \rangle}\bigg|_{R_0}}{\Delta E_{\rm NS}(ns,R_0)}=\frac{\frac{d \Delta E_{\rm NS}(np_{1/2},R)}{d \langle r^2 \rangle}\bigg|_{R_0}}{\Delta E_{\rm NS}(np_{1/2},R_0)}=\frac{\gamma}{R_0^2},
\end{equation}
where $\gamma=\sqrt{1-(\alpha Z)^2}$. 
Our numerical calculations within the \textsc{hfd} code~\cite{HFD,HFDB} confirm this analytical result with high accuracy. The second term in the square brackets in Eq. (\ref{SEderiv}) is independent of $n$ for $ns$ and $np_{3/2}$ states according to the explicit expression for 
$G_{\rm NSE}(nlj,R)$ 
given in Ref.~\cite{Yerokhin:2011b}. A certain dependence on $n$ does exist for $np_{1/2}$ states due to the dependence of the parameter $a_{21}=-2(n^2-1)/n^2$, which enters into the expression for 
$G_{\rm NSE}(nlj,R)$ 
(see Eq.~(11) of Ref.~\cite{Yerokhin:2011b}). To explore this dependence, we set $\sqrt{\langle r^2 \rangle} \approx (2Z)^{1/3} + 0.57$ fm approximately following the empirical dependence of $\sqrt{\langle r^2 \rangle}$ on the nucleon number ~\cite{Johnson:1985}. We found that non-negligible dependence occurs in the region of $Z=36-40$. If we are outside this range, one can write:
\begin{equation}
\label{SEderiv2}
   \frac{d \Delta E^{\rm SE}(nlj,R)}{d \langle r^2 \rangle} \bigg|_{R_0} =
    \Delta E^{\rm SE}(nlj,R_0) M(lj,R_0),   
\end{equation}   
where $M(lj,R)$ is a function of $l,j,R$, but not $n$. 
Putting it all together, we obtain that
\begin{equation}
\label{deriv}
    \frac{d X_{kljm,k'ljm}(\langle r^2 \rangle)}
    {d \langle r^2 \rangle} = X^{\rm NS}_{kljm,k'ljm} M(lj,R_0),
\end{equation}
where $X^{\rm NS}_{kljm,k'ljm} = X_{kljm,k'ljm}\bigg|_{R_0} - X_{kljm,k'ljm}\bigg|_{0}$ represents the NS correction to the matrix element $X_{kljm,k'ljm}$. The derivatives in Eq.~(\ref{deriv}) can be readily evaluated as the nuclear size contribution to all required diagonal and off-diagonal matrix elements of SE are available from Ref.~\cite{Shabaev:13}.
To cover the range of $Z$ mentioned above, a more accurate expression, which takes into account the principal quantum number dependence in the calculation of SE matrix elements over the H-like $p_{1/2}$ functions, is to be considered, but this is out of the scope of the present interest.
Note that the first term in Eq. (\ref{Fse}) can be calculated as an expectation value of the operator
\begin{equation}
    F^{\rm SE,1} \approx \sum_{k,k',ljm} |h_{kljm}\rangle X'_{kljm,k'ljm} \langle h_{k'ljm}|,
\end{equation}
where $X'_{kljm,k'ljm}=\frac{d X_{kljm,k'ljm}(\langle r^2 \rangle)}{d \langle r^2 \rangle}$.

\begin{table*}
\caption{QED contributions to the field shift $F$ constant (in MHz/fm$^2$) for the $2p_{1/2} \to 2s_{1/2}$ transition in Li-like ions. }
\label{TQEDcomp}
\begin{tabular*}{1.0\textwidth}{l@{\extracolsep{\fill}}llllll}
\hline
\hline
Ion                    & VP(Uehling)           & SE       &\textbf{Total, this work} & Zubova et. al, 2014 \cite{Zubova:2014} & Zubova et. al, 2016~\cite{Zubova:2016} \\
\hline
Ar$^{15+}$ & -24                 & 64                  & 40                  & 44                       & --                        \\
Bi$^{78+}$ & -0.099$\times 10^7$ & 0.137$\times 10^7$  & 0.038$\times 10^7$  & 0.039(11) $\times 10^7$  & 0.0439(35) $\times 10^7$    \\
U$^{87+}$  & -0.276$\times 10^7$ & 0.385$\times 10^7$  & 0.109$\times 10^7$  & 0.087(30) $\times 10^7$  & 0.1026(82) $\times 10^7$    \\
\hline
\hline
\end{tabular*}
\end{table*}

In Ref.~\cite{Shabaev:13}, the SE matrix elements for the point nucleus as well as the NS corrections for these elements are provided for many elements with step in $Z$ equal to 5. For the other elements, it is proposed to use an interpolation formula suggested in Ref.~\cite{Mohr:83}. However, in the present case, we are studying a more delicate problem than the total SE matrix elements. For the approach introduced above, the finite nuclear size SE effect is important. To calculate the corresponding matrix elements, one can compute the difference between the interpolated SE matrix elements with and without including the finite nuclear size effect. However, there is some inconsistency in such an approach, as the finite nuclear size SE corrections tabulated in Ref.~\cite{Shabaev:13} were obtained for the specific rms charge radii of reference elements (e.g. $Z$=10,15,20, etc). According to our analysis, a more stable interpolation result for the NS contribution to SE can be obtained if one first scales the NS SE contributions for the reference elements included in the interpolation procedure to the required rms charge radius of an element under consideration. For such a scale, one can use Eq.~(\ref{Ederiv}). The reliability of the described scheme was verified by performing \textit{ab initio} calculations of the NS SE contributions directly for Al (see below).

The approach that we introduced above can be combined with methods to treat high-order electron correlation effects in many-electron atoms and even in molecules, such as the coupled cluster method with single, double, triple, and perturative quadruple excitations CCSDT(Q)~\cite{Kallay:6,Bartlett:2007}, which includes all terms of the perturbation theory up to the sixth order and some terms to all orders. The simultaneous consideration of QED and electron correlation effects is important for such systems as discussed below. We should mention that in Ref.~\cite{King:2022:43} probably a similar strategy of estimating the SE contribution to the FS constant was applied to boron-like Ar, although without a discussion and details.

\subsection{\label{sec:recoilQED} QED contributions to the mass shift}
The rigorous QED theory of the nuclear recoil effect beyond the Breit approximation has been developed in Ref.~\cite{shabaev1985mass, shabaev1988nucl, ShabaevRecoil:98}, see also Refs.~\cite{Pachucki:1995:1854, Yelkhovsky:Budker, Adkins:2007:042508}. In the leading order it reduces to the approximation given by Eqs.~(\ref{opNMS}) and (\ref{opSMS}). Recently, a model QED approach has been formulated to account for the QED corrections to the one electron part of the mass shift~\cite{Anisimova:2022} beyond the approximation (\ref{opNMS}). In this reference, the QED effects on the nuclear recoil were encoded into the effective operator that has the same structure as those introduced in Ref.~\cite{Shabaev:13} for the SE effect. Both diagonal and off-diagonal matrix elements over the H-like functions were calculated and tabulated for many elements, and corresponding interpolation formulas were provided where necessary. Some test applications have been outlined~\cite{Anisimova:2022}.
In the present paper, we have adapted this approach by replacing the SE matrix elements with the nuclear recoil QED matrix elements in the operator (\ref{Xmod}), which has already been interfaced~\cite{Skripnikov:2021a} to relativistic program packages, allowing one to perform correlation calculations for many-electron systems.

There is an additional QED contribution to the nuclear recoil effect. It is induced by the perturbation of the electronic wave function by the SE and VP interactions~\cite{King:2022:43,Sahoo:2021}. This effect can be taken into account by calculating the difference between the values of the mass shift constants calculated with and without inclusion of the Uehling and model SE operators into the electron correlation calculation.

\section{Calculations}
In this section, we perform test calculations of the QED effects, provide details of the electronic structure calculations of the IS atomic factors in Al including a discussion of different approaches for their calculation, and finally outline the uncertainty estimation procedure.

\subsection{Test of model QED calculations}
In the previous section, we introduced a model approach to calculate the QED corrections to the FS constant $F$. The most straightforward way to validate such an approach is to compute the QED corrections for $F$ for highly charged ions, for which rigorous \textit{ab-initio} QED calculations are available and where a clear comparison can be made. This is particularly feasible when electronic correlation effects are suppressed, allowing one to focus on the QED effect of interest. In Ref.~\cite{Zubova:2016}, such calculations were performed for Li-like ions.

Table \ref{TQEDcomp} gives the values of the QED contributions to the FS constant of the $2p_{1/2} \to 2s_{1/2}$ transition in Li-like HCIs. Electron correlation effects were considered within the relativistic coupled cluster with single and double excitation amplitudes method, CCSD. For the elements considered, we used the AAE4Z basis sets~\cite{Dyall:2016,Dyall:12,Dyall:07} truncated after $d$-type functions and the Dirac-Coulomb-Breit Hamiltonian.

One can see an agreement within 15\% between the QED contributions to the FS constants in Li-like Bi and U obtained in the present work and within the rigorous QED treatment in Ref.~\cite{Zubova:2016}. For Ar, we also observe a good agreement with the QED estimation performed in Ref.~\cite{Zubova:2014}. The small deviation may be caused by some higher-order QED terms considered in Ref.~\cite{Zubova:2016} but omitted in our model approach.The difference between our values and that of Ref.~\cite{Zubova:2016} can be considered as an uncertainty estimation of the present approach. The achieved agreement is already more than sufficient for the present purposes and surpasses previous attempts of the effective QED calculations~\cite{Sahoo:2021}. It's worth noting that in the Al atom of the present interest, we also consider the $s \to p$ type transitions, much like in these Li-like HCIs.

In section \ref{sec:recoilQED} we also adapted the model approach to the nuclear recoil effect beyond the approximation given by the operator~(\ref{opNMS}), that is, to the corresponding QED contribution. To test the approach, we calculated this QED contribution to the mass shift constant for the $2p_{1/2} \to 2s_{1/2}$ transition in Li-like U, for which rigorous QED calculations are available~\cite{MalyshevQEDNMS:2020}. In the independent particle approximation, our calculated QED contribution, $-2221921$ GHz u, practically coincides with that of Ref.~\cite{MalyshevQEDNMS:2020}, $-2221736$ GHz u, as expected due to the model QED operator definition. More interestingly, the correlation contribution calculated in this work, 60973 GHz u, is also in excellent agreement with the rigorous QED result~\cite{MalyshevQEDNMS:2020}, 59543 GHz u.

\subsection{Computational details}

To achieve a balanced description of relativistic and electronic correlation effects, we employed the relativistic coupled cluster method with single, double, and iterative full triple excitations, CCSDT, method for all electrons of Al and used the Dirac-Coulomb Hamiltonian \cite{Bartlett:2007}. Note, that all triple excitations from all the electronic shells were considered. In the correlation calculations, the virtual energy cutoff was set to 500 Hartree. We used the uncontracted augmented all-electron quadruple-zeta AAE4Z basis set~\cite{Dyall:2016}, extended by additional $s$-, $p$-, and $h$-type functions. In total, this basis set includes $31s21p10d7f4g4h$ Gaussian-type functions and will be referred to as MBas below.

To account for the higher-order correlation effects, we calculated the difference between the values of the considered properties obtained within the relativistic CCSDT(Q)~\cite{Kallay:6}, and the CCSDT methods. These calculations were performed within the uncontracted augmented all-electron triple-zeta AAE3Z basis set \cite{Dyall:2016}, extended by additional $s$-, $p$-, and $d$-type functions, and will be referred to as SBas. About $2 \times 10^{10}$ quadruple excitations were taken into account perturbatively.

To account for the effects of a larger basis set, the following corrections were computed. (i) First, we evaluated the difference between the values obtained within the CCSD method using a very large basis set, LBas, and the basis MBas used in our main model. The LBas basis set includes $50s50p33d20f9g7h6i$ functions. For $s$- and $p$- type functions, we used the exponent parameters forming a geometric progression with a common ratio of 1.77, and the largest term was set to $7\times 10^8$. For d-type functions, we used a progression with a common ratio of 1.85, and the largest term was set to $1\times 10^6$. For f-type functions, a progression with a common ratio of 1.8 and the largest term equal to 500 was used. For $g$-, $h$-, and $i$-type functions, we used the natural-like basis set constructed within the procedure developed in Refs. \cite{Skripnikov:2020e,Skripnikov:13a}. (ii) Second, we calculated the contribution of $k$-type functions ($L=7$). Finally we computed the extrapolated contribution of basis functions with higher $L$ using the approach described in Refs. \cite{Skripnikov:2021a,AthanasakisRaFPinning:2023}.

Next, we calculated the correlation contribution induced by including virtual orbitals above 500 Hartree. This correction was computed using the relativistic coupled cluster with single, double, and perturbative triple excitations, CCSD(T), method \cite{Visscher:96a,Bartlett:2007} for all properties except for the SMS constant, for which this correction was calculated at the CCSD level.

At the next stage, we calculated the contribution of the Breit interelectronic interaction in the zero-frequency limit. For this, we computed the difference between the values obtained within the Dirac-Coulomb-Breit and Dirac-Coulomb Hamiltonians at the CCSDT level using the SBas basis set and without using the virtual energy cutoff.

Finally, we calculated the contributions of the QED effects. For transition energies, we used the model SE and VP approach~\cite{Shabaev:13,Skripnikov:2021a}. To compute the QED contribution to the nuclear recoil effect, we employed the model operator developed in Ref.~\cite{Anisimova:2022}, which was adapted for the high-order correlation calculations in the present paper. We also took into account the QED contribution due to perturbation of the electron wave function by the model SE and VP operators. To account for the contribution of the QED effects on the FS constant, we employed the approach developed in the present work.

Relativistic four-component calculations were performed within the locally modified {\sc dirac15} code \cite{DIRAC15,Saue:2020}. The basis set corrections for the SMS factor at the CCSD level were calculated using the {\sc exp-t} code \cite{EXPT_website,Oleynichenko_EXPT,Zaitsevskii:2023}. The high-order correlation effects within the CCSDT, CCSDT(Q), the coupled cluster single, double, triple, and quadruple, CCSDTQ and the coupled cluster single, double, triple, quadruple and perturbative 5-fold excitations CCSDTQ(P)~\cite{Kallay:6,Bartlett:2007}, approaches were computed using the locally modified {\sc mrcc} code \cite{MRCC2020}. To treat the nuclear Fermi charge distribution, we used the code developed in Ref.~\cite{Skripnikov:2024a}. Matrix elements of the NMS and SMS operators were calculated using the code developed in Ref.~\cite{Penyazkov:2023}. Two-electron integrals of the Breit interaction operator over atomic bispinors were computed using the code from Ref.~\cite{Maison:2019}.

\subsection{Property calculations}

While calculations of the transition energies are straightforward, questions often arise about how to calculate properties such as the IS  factors~\cite{Sahoo_2020,Sahoo:2021}. In experiments, the IS is obtained by comparing transition energies for two isotopes, which have different ms charge radii. In theoretical treatments one can use this approach by computing the transition energies for a given atom and two different ms charge radii. Then, the FS constant can be determined as the ratio of the computed energy shift to the shift of the ms charge radii used in the calculation. In Eq.~(\ref{FSdef}), one should perform this calculation with an infinitesimal ms shift. Alternatively, the field shift constant for a given state can be calculated using the following expression:
\begin{equation}
\label{FSexp}
   \langle O \rangle =\langle \Psi | \sum_i O(i) | \Psi \rangle,
\end{equation}
where $O=d H_{\rm nuc}(\langle r^2 \rangle) / d \langle r^2 \rangle$, $H_{\rm nuc}$ is the electron-nucleus interaction operator, and $\Psi$ is the many-electron wave function. In other words, one can calculate the FS as an expectation value of the one-electron operator $d H_{\rm nuc} /d \langle r^2 \rangle$.  There are also different possibilities how one can calculate the expectation value. For example, one can perform an analytical calculation of the one-particle density matrix and evaluate the trace of this matrix with the corresponding matrix elements. On the other hand, one can calculate it using the finite field technique. According to the Hellmann–Feynman theorem, for the exact electronic wave function the results calculated within the different approaches should be the same. We performed test calculations using three strategies to evaluate the FS constant. 

(i) In the method \textbf{A-relaxed} we calculated 
\begin{equation}
\label{Frelax}
F^{\rm (A,relaxed)}=\frac{E(\langle r_0^2\rangle+h) - E(\langle r_0^2\rangle-h)}{2h}    
\end{equation}
Here $\langle r_0^2\rangle=3.061^2$~fm$^2$ is the ms charge radius of $^{27}$Al~\cite{ANGELI201369,Heylen:2021} and $h=0.5$~fm$^2$. In this calculation we directly changed the nuclear potential operator corresponding to the Gaussian nuclear charge distribution. This change has been made already at the DHF stage of the calculation. In the present work, we consider two specific models of the nuclear charge distribution: the Gaussian and Fermi ones (see below). When we change the rms charge radius, other nuclear moments also change according to the considered model of the charge distribution. This should be taken into account when considering Eq.~(\ref{FSdef}).

(ii) In the method \textbf{A-unrelaxed} the field shift constant was calculated as 
\begin{equation}
 F^{\rm (A,unrelaxed)}=\frac{E(\lambda) - E(-\lambda)}{2\lambda},
\end{equation}
where $E(\lambda)$ is the total energy calculated for the electronic Hamiltonian perturbed by the operator $(H_{\rm nuc}(\langle r^2 \rangle + h) - H_{\rm nuc}(\langle r^2 \rangle - h))/2h$ with a factor $\lambda$. In calculations we set $\lambda=1$. The perturbating operator was added to the Hamiltonian \textit{after} the DHF stage of the calculation, but before the correlation stage. Therefore, the orbital relaxation effects were not considered. One-electron orbitals were obtained for the reference isotope with the ms charge radius $\langle r_0^2\rangle$.

(iii) Finally, in method \textbf{B}, we performed calculations corresponding to the expression (\ref{FSexp}). Namely, we calculated the one-particle density matrix $D$ using the $\Lambda$-equations coupled cluster technique~\cite{Kallay:3,Bartlett:2007}, and then evaluated the trace of this matrix with the matrix elements of the operator 
$d H_{\text{nuc}}/d \langle r^2 \rangle$
\begin{equation}
 F^{\rm (B)}=\sum_{p,q}D_{p,q}\left(\frac{d H_{\rm nuc}}{d \langle r^2 \rangle}\right)_{p,q}.
\end{equation}
In all cases we used the Gaussian nuclear charge distribution model
~\cite{Visscher:1997} with the Gaussian nuclear charge distribution parameter 
$\xi=1.5/\langle r^2 \rangle$,
for which the $H_{\rm nuc}$ operator is
\begin{equation}
   H_{\rm nuc}=-\frac{Z}{r}{\rm erf}(\sqrt \xi r).
\end{equation}
In this case, the derivative $d H_{\text{nuc}}/d \langle r^2 \rangle$
can be taken analytically:
\begin{equation}
   \frac{d H_{\text{nuc}}}{d \langle r^2 \rangle}=\frac{2Z \xi^{3/2}}{3\sqrt \pi} e^{-\xi r^2}.
\end{equation}
Thus, in contrast to the A-relaxed and A-unrelaxed methods, the method B does not suffer from the numerical differentiation errors.

In all these test calculations, we used the manually extended uncontracted all-electron double-zeta AE2Z basis set~\cite{Dyall:2016} and different levels of electronic correlation effects treatment. This basis set is smaller than that used in the actual calculation scheme described above. However, it allowed us to consider the correlation effects up to the relativistic CCSDTQ(P) method. No virtual orbitals energy cutoff was applied, and all electrons were included in the correlation treatment. Table \ref{FScompare} summarizes the results of applying three strategies of calculating the IS factor for the $3s^23p~^2P_{1/2} \to 3s^24s~^2S_{1/2}$ transition in the Al atom.
\begin{table}[]
\caption{Values of the field shift constant $F$ (in MHz/fm$^2$) for the $3s^23p~^2P_{1/2} \to 3s^24s~^2S_{1/2}$ transition in Al within the different strategies to calculate this property and different methods to treat the electron correlation effects. A small basis set was employed for these test calculations.}
\begin{tabular}{l
                S[table-format=5.3,group-separator=,table-align-text-post=false]
                S[table-format=5.3,group-separator=,table-align-text-post=false]
                S[table-format=5.3,group-separator=,table-align-text-post=false]}
\hline
\hline
Method   & {A-relaxed} & {~A-unrelaxed} & {B}    \\
\hline         
DHF      & 63.69     & 13.20       & 13.20 \\
CCSD     & 74.46     & 74.62       & 74.62 \\
CCSD(T)  & 74.09     & 73.92       &       \\
CCSDT    & 73.81     & 73.82       & 73.82 \\
CCSDT(Q) & 73.79     & 73.80       &       \\
CCSDTQ   & 73.79     & 73.79       &       \\
CCSDTQ(P) & 73.79     & 73.79       &       \\
\hline
\hline
\end{tabular}
\label{FScompare}
\end{table}

Note that the $\Lambda$-equations approach used in the method B is not implemented for the CCSD(T), CCSDT(Q) and CCSDTQ(P) models, while for the CCSDTQ model it was found to be too computationally expensive. As $h \to 0$, $[H_{\text{nuc}}(\langle r^2 \rangle + h) - H_{\text{nuc}}(\langle r^2 \rangle - h)] / 2h \to d H_{\text{nuc}}/d \langle r^2 \rangle$,
and in the absence of numerical differentiation errors, this model becomes identical to the A-unrelaxed model. As one can see from Table~\ref{FScompare}, we indeed have good agreement between the predictions of these models. Therefore, the numerical differentiation errors, which can potentially arise in the A-unrelaxed method, are negligible at the considered level of precision within the considered single-reference CC methods (similar conclusion was also made for other properties~\cite{Skripnikov:2024a,Skripnikov:2020b}).

The methods A-relaxed and A-unrelaxed mainly differ in the way they account for the orbital relaxation effects. However, it is expected that for the exact treatment of the electron correlation effects, both methods should yield identical results. The deviation between the results can provide partial information about the quality of the nonvariational method used to treat the electron correlation effects. However, the coincidence of the results does not necessarily mean that the correlation is accounted for exactly. Table \ref{FScompare} compares the results obtained within these two methods. We begin with the DHF values, which differ by a factor of 5. However, already at the CCSD level, the deviation is 0.22\%. It slightly increases at the CCSD(T) level to 0.24\%. However, at the higher-order correlation treatment levels, the methods A-relaxed and A-unrelaxed give essentially the same results with deviation smaller than 0.01\%.

Let us now perform a similar analysis for the contribution of the first term of the normal mass shift effect (\ref{opNMS}). As one can see, this operator is proportional to the nonrelativistic operator of the kinetic energy $T$. Similar to the FS case, we can consider three methods of calculating this effect in the first order.
(i) In the A-relaxed method, we added this operator at the DHF stage of electronic calculation to account for the orbital relaxation effects. Then, the effect was calculated as the numerical derivative 
$[E(\lambda) - E(-\lambda)]/2\lambda$, 
where $E(\lambda)$ is the total energy value obtained when the operator $T$ is added to the electronic Hamiltonian with the factor $\lambda=10^{-5}$. 
(ii) In the method A-unrelaxed, the operator was added after the DHF stage, but before the coupled cluster stage. The value was also calculated as the numerical derivative using the same value of the $\lambda$ parameter. 
(iii) In the method B, we calculated the trace of the one-particle density matrix with the matrix elements of the operator~$T$. No numerical derivatives were needed in this case.

Table \ref{NNMScompare} presents the results obtained from the application of three methods for calculating the contribution of the first term in Eq.~(\ref{opNMS}) to the NMS atomic factor for the $3s^23p~^2P_{1/2} \to 3s^24s~^2S_{1/2}$ transition in aluminium. The computational details are identical to those described above for the FS.

\begin{table}[]
\caption{
Values of the first term of Eq.~(\ref{opNMS}) to the NMS effect (in GHz u) for the $3s^23p~^2P_{1/2} \to 3s^24s~^2S_{1/2}$ transition in Al within the different strategies to calculate this property and different methods to treat the electron correlation effects. A small basis set was employed for these test calculations.}
\begin{tabular}{l
                S[table-format=6.3,group-separator=,table-align-text-post=false]
                S[table-format=7.3,group-separator=,table-align-text-post=false]
                S[table-format=6.3,group-separator=,table-align-text-post=false]}
\hline
\hline
Method   & {A-relaxed} & {~A-unrelaxed} & {B}          \\
\hline
DHF       & -356.79    & -1487.44    & -1487.44 \\
CCSD     & -400.26    & -391.35      & -391.35  \\
CCSD(T)  & -399.68    & -403.61      &          \\
CCSDT    & -399.25    & -399.32      &  -399.32 \\
CCSDT(Q) & -399.23    & -399.61      &          \\
CCSDTQ   & -399.24    & -399.24      &          \\
CCSDTQ(P)  & -399.24    & -399.25      &          \\
\hline
\hline
\end{tabular}
\label{NNMScompare}
\end{table}

As in the FS constant calculation, in the present case we again obtain results within the schemes A-unrelaxed and B that, as expected, deviate negligibly at all levels of correlation treatment. Next, we compare the A-relaxed and A-unrelaxed results. At the CCSD level, the deviation between these methods is approximately 2\%. As we progress to the CCSD(T) level, the deviation reduces to less than 1\%, and at higher levels of correlation treatment the results almost coincide, as expected for such high-level models as the full iterative CCSDT approach or higher. The A-relaxed method demonstrates a faster convergence for the considered property (see also  Ref.~\cite{Skripnikov:17a} with a related discussion about the strategy of calculating the atomic enhancement factor of the scalar-pseudoscalar nucleus-electron interaction).

Our comparison of the strategies of the property calculations was performed for the single-reference coupled cluster methods employed in the present work. Different methods of handling electron correlation effects may behave differently.

In our study of the IS in Al, we used the A-relaxed scheme to calculate the FS constant and the first term of Eq.~(\ref{opNMS}) for the NMS effect. 
For the remaining terms of the nuclear recoil effects, we applied the A-unrelaxed approach due to technical reasons. It's worth highlighting that employing a hierarchy of single-reference CC methods enables a systematic consideration of uncertainties.

\begin{table*}[]
\caption{Values of the transition energies in Al (in cm$^{-1}$).}
\begin{tabular}{l
                S[table-format=-3.2(2), table-align-text-post=false, table-space-text-post=]
                S[table-format=-3.2(2), table-align-text-post=false, table-space-text-post=]
                S[table-format=-3.2(2), table-align-text-post=false, table-space-text-post=]
                S[table-format=-3.2(2), table-align-text-post=false, table-space-text-post=]}
\hline
\hline
                 & {$3s^23p~^2P_{1/2} \to 3s^24s~^2S_{1/2}$} & {$3s^23p~^2P_{3/2} \to 3s^24s~^2S_{1/2}$}  & {$3s^23p~^2P_{1/2} \to 3s^25s~^2S_{1/2}$}    & {$3s^23p~^2P_{3/2} \to 3s^25s~^2S_{1/2}$}  \\
\hline                 
CCSDT              & 25315     & 25197     & 37644     & 37526     \\
CCSDT(Q) $-$ CCSDT & 0         & 1         & 7         & 9         \\
basis set corr.    & 28        & 28        & 32        & 32        \\
high virt.         & -1        & -1        & -1        & -1        \\
Breit              & -8        & 0         & -8        & -1        \\
QED                & 5         & 5         & 5         & 4         \\
Total, this work                    & 25339(10)  & 25230(9)   & 37679(13)     & 37570(14)     \\
\\
MCDHF(CV) Ref.~\cite{Filippin:2016} & 25495      &  25376   &   &  \\
MCDHF(CC) Ref.~\cite{Filippin:2016} & 25351      &  25173   &   &  \\
Experiment~\cite{AlEEKaufman:1991}  & 25347.756  & 25235.695 & 37689.407 & 37577.346 \\
\hline
\hline
\end{tabular}
\label{Tenergies}
\end{table*}

\subsection{Uncertainty estimation}
We considered the following sources of theoretical uncertainties for the calculated IS atomic factors and transition energies in Al:

(i) The uncertainty stemming from the unaccounted correlation effects beyond the CCSDT(Q) model was estimated by considering the contribution of perturbative connected quadruple excitation amplitudes, specifically the difference between the CCSDT(Q) and CCSDT results.

(ii) To address the uncertainty due to the basis set incompleteness, we followed a two-stage approach. First, we compared the basis set corrections to the electron transition energies, FS, and NMS factors calculated at the CCSD(T) and CCSD levels. This comparison allowed us to estimate the uncertainty arising from the reduced level of the electron correlation effects treatment for the basis set correction calculation. For the SMS factor, such an estimation was not feasible due to technical reasons, so the total value of the basis set correction was used. Second, our calculation scheme involved the basis functions up to $L=7$. We used the doubled value of the $k$-type function contribution as an estimation of possible higher-order harmonics contribution to the considered IS parameters and the sum of the $k$- type function contribution and an extrapolated higher-order harmonics contribution for the case of transition energies.

(iii) The uncertainty associated with the contribution of higher-lying virtual orbitals (with the orbital energies exceeding 500 Hartree) was evaluated by comparing this contribution calculated in two basis sets, MBas and SBas, at the CCSD(T) levels. It's noteworthy that the corrections calculated within the CCSD and CCSD(T) levels nearly coincide with each other.

(iv) For the uncertainty of the QED effects, we conservatively set it to 30\% of their values.

(v) As a measure of the uncertainty due to the missed $\omega$-dependence in the Breit electron-electron interaction we conservatively set it to 50\% of the calculated Breit correction.

(vi) Concerning the FS constant, we also compared the different models of the nuclear charge distribution: the Gaussian and Fermi ones. The discrepancy in the results obtained within these two models was included in the uncertainty of the FS constant (see also the analysis below).

The total uncertainty was estimated as the square root of the sum of the squares of all the aforementioned uncertainties.

\section{Results and discussion}

The calculated values of the excitation energies in neutral Al for the electronic states of interest are given in Table~\ref{Tenergies}. One can see good agreement between calculated and experimental values~\cite{AlEEKaufman:1991}. 
%

The calculated values of the atomic IS factors are given in Table~\ref{AlISfactors}. 
\begin{table*}[]
\caption{Values of the IS atomic factors for Al.}
\begin{tabular}{l
                S[table-format=-3.2(2), table-align-text-post=false, table-space-text-post=]
                S[table-format=-3.2(2), table-align-text-post=false, table-space-text-post=]
                S[table-format=-3.2(2), table-align-text-post=false, table-space-text-post=]
                S[table-format=-3.2(2), table-align-text-post=false, table-space-text-post=]}
\hline
\hline
\multicolumn{5}{c}{$k_{\rm NMS}$, GHz u}                                                          \\
                 & {$3s^23p~^2P_{1/2} \to 3s^24s~^2S_{1/2}$} & {$3s^23p~^2P_{3/2} \to 3s^24s~^2S_{1/2}$}  & {$3s^23p~^2P_{1/2} \to 3s^25s~^2S_{1/2}$}    & {$3s^23p~^2P_{3/2} \to 3s^25s~^2S_{1/2}$}  \\
\hline                 
CCSDT              & -415.4 & -414.4 & -618.1 & -617.0 \\
CCSDT(Q) $-$ CCSDT & 0.0    & 0.0    & -0.1   & -0.1   \\
basis set corr.    & -0.7   & -0.7   & -0.8   & -0.8   \\
high virt.         & 0.1    & 0.1    & 0.1    & 0.1    \\
Breit              & 0.4    & 0.1    & 0.4    & 0.1    \\
QED                & 1.0    & 1.0    & 0.9    & 0.9    \\
Total              & -414.7$(0.5)$ & -413.9$(0.3)$ & -617.6$(0.5)$ & -616.8$(0.4)$ \\
                   &             &             &             &           \\
\multicolumn{5}{c}{$k_{\rm SMS}$, GHz u}                                                          \\
\hline
CCSDT              & 653.4 & 655.1 & 617.9 & 619.6 \\
CCSDT(Q) $-$ CCSDT & -0.1  & -0.6  & -1.9  & -2.3  \\
basis set corr.    & 0.8   & 0.8   & 0.8   & 0.8   \\
high virt.         & 0.5   & 0.5   & 0.4   & 0.4   \\
Breit              & -0.5  & 0.0   & -0.5  & 0.0   \\                
QED                & 0.2    & 0.2    & 0.2    & 0.2    \\
Total              & 654.3$(0.9)$ & 656.0$(1.0)$ & 616.9$(2.1)$ & 618.6$(2.5)$ \\
                   &                      &            &              &            \\
\multicolumn{5}{c}{$k_{\rm NMS} + k_{\rm SMS}$, GHz u   }   \\
\hline
Total, this work   & 239.6$(1.1)$   &  242.1$(1.0)$  & -0.7$ (2.1)$   &   1.7$(2.5)$     \\
MCDF(CV+VV)~\cite{Filippin:2016,Heylen:2021}  & 240.0$(5.0)$ $^a$ &  243.0$(4.0)$    &              &               \\
\hline                   
                   &                      &            &              &            \\
\multicolumn{5}{c}{ $F$, MHz/fm$^2$   }                                                          \\
\hline
CCSDT              & 76.95 & 76.85 & 70.15 & 70.05 \\
CCSDT(Q) $-$ CCSDT & -0.02 & -0.01 & 0.02  & 0.03  \\
basis set corr.    & 0.04  & 0.05  & 0.07  & 0.08  \\
high virt.         & 0.01  & 0.01  & 0.01  & 0.01  \\
Breit              & -0.07 & 0.00  & -0.05 & 0.02  \\
QED-VP             & 0.07  & 0.06  & 0.06  & 0.06  \\
QED-SE             & -0.17 & -0.17 & -0.15 & -0.15 \\
Total, this work              & 76.81(12) & 76.80(6) & 70.11(13) & 70.10(8) \\
Refs.~\cite{Filippin:2016,Heylen:2021}    &   76.45$(1.95)$$^a$        & 76.20$(2.20)$  &              &            \\
\hline
\hline
\end{tabular}
\begin{flushleft}
$^a$ Derived from the data given in Table III of Ref.~\cite{Filippin:2016} (columns RIS3/Sep. and RATIP/Sep.) following the approach described in Ref.~\cite{Heylen:2021}, see Refs.~\cite{Filippin:2016,Heylen:2021} for the description and abbreviations of these methods.
\end{flushleft}
\label{AlISfactors}
\end{table*}
The uncertainty estimation procedure is described in the previous section. Note that the exact nuclear charge distribution for the Al isotopes is unknown. In our calculation, we used the one-parameter Gaussian model of the nuclear charge distribution~\cite{Visscher:1997}. However, considering the similar models of the charge distributions for a pair of isotopes may be too restrictive~\cite{Blundell:2018}. To test a possible change in the nuclear shape, we compared the values of the FS constants calculated within two approaches using Eq.~(\ref{Frelax}). In the first approach, we calculated the numerical derivative (\ref{Frelax}) using the Gaussian nuclear charge distribution for both terms in Eq.~(\ref{Frelax}). In the second approach, the term with the ms charge radius $\langle r_0^2\rangle+h$ was calculated within the Gaussian nuclear charge distribution, while the term with the ms charge radius $\langle r_0^2\rangle-h$ was calculated within the Fermi nuclear charge distribution. Although the Gaussian and Fermi distributions differ very significantly~\cite{maartensson2003atomic}, we found that the values of the FS constants calculated within the first and second approaches differ by less than 0.007 MHz/fm$^2$ for all of the considered transitions. This effect is negligible with respect to other uncertainties. However, to be conservative, we included the value 0.01 MHz/fm$^2$ in the uncertainty estimation of the FS constant to account for the effect of the unknown nuclear shape (see item (vi) in the previous section).

It is evident that the small uncertainties in the non-QED calculations of the atomic IS factors necessitate the consideration of the QED effects. These effects are typically overlooked in theoretical IS studies of many-electron neutral atoms due to uncertainties arising from the incomplete treatment of the correlation effects. Of particular interest is the SE QED contribution $F^{\rm SE}$ to the FS constant, $F^{\rm SE}$ which was found to be larger than the VP contribution (compare lines ``QED-SE'' and ``QED-VP'' of Table \ref{AlISfactors}). In Section \ref{SecFSqed}, we proposed an effective method for performing such calculations for many-electron systems, which allows the use of the relativistic coupled cluster methods. The challenge arises from the lack of exact knowledge of the contribution of the finite nuclear size effect $\Delta E^{\rm SE}(nlj,R_0)$ in Eq.~(\ref{SEderiv2}). While the literature provides calculations for $Z=10$, 15, 20, etc.~\cite{Shabaev:13}, it does not cover $Z=13$. In the section~\ref{SecFSqed} we mentioned that the interpolation procedure can be used. The good agreement between the proposed technique and the rigorous QED treatment, as shown in Table~\ref{TQEDcomp} for Li-like ions, suggests that interpolation is viable. However, since Al is a light atom with $Z=13$, the interpolation scheme for it is based on only one point with a smaller $Z$ ($Z=10$). To verify the reliability of our approach, three calculations of the field shift constant were performed for the  $3s^23p~^2P_{1/2} \to 3s^24s~^2S_{1/2}$ transition in Al:
(i) employing Eqs.~(\ref{Fse}), (\ref{deriv}) and the matrix elements $X^{\rm NS}_{kljm,k'ljm}$ interpolated using the data from Ref.~\cite{Shabaev:13}, along with the analytical expressions for the derivative coefficients $M(lj,R_0)$ obtained with the use of Eq.~(\ref{SEderiv});
(ii) using Eq.~(\ref{Fse}) with the directly calculated finite nuclear size contributions to the SE matrix elements and numerically computed 
$d X_{p,q}^{\langle d r^2 \rangle} / d \langle r^2 \rangle$ 
matrix elements within the \textit{ab initio} QED methods~\cite{Yerokhin:1999:800, Malyshev:2022, Sapirstein:2023:042804};
(iii) directly applying Eq.~(\ref{FQED}). In the latter case, two sets of the SE matrix elements over the  H-like functions were specially calculated for Al for two different nuclear ms charge radii. Consequently, $F^{\rm SE}$ was calculated directly as a numerical derivative.
The approach (iii) is the most accurate, while the approach (i) is the most approximate but relies solely on the QED data published in the seminal paper~\cite{Shabaev:13} and does not require any new QED calculations. This significantly simplifies the problem of calculating the QED contribution, since an accurate evaluation of the SE matrix elements is not a trivial issue.

\begin{table}[h]
\caption{Values of the first-type QED contribution (see main text) to the FS constant, $F^{\rm SE}$, for the  $3s^23p~^2P_{1/2} \to 3s^24s~^2S_{1/2}$ transition in Al obtained using three approaches (see main text) at different levels of electron correlation treatment (in MHz/fm$^2$).}
\label{TabFqedAl}
\resizebox{\columnwidth}{!}{%
\begin{tabular}{l
                S[table-format=-1.3, table-align-text-post=false]
                S[table-format=-1.3, table-align-text-post=false]
                S[table-format=-1.3, table-align-text-post=false]}
\hline
\hline
                                                 & {DHF}    & {CCSD}   & {CCSD(T)} \\
\hline                                                                         
Term 1 of Eq.~(\ref{Fse}), extrap., analyt. & -0.031 & -0.182 & -0.181  \\
Term 2 of Eq.~(\ref{Fse}), extrap., analyt.  & 0.001  & 0.010  & 0.009   \\
Sum                                               & -0.030 & -0.172 & -0.171  \\
                                                  &        &        &         \\
Term1 of Eq.~(\ref{Fse}), numerical calc.         & -0.032 & -0.183 & -0.181  \\
Term2 of Eq.~(\ref{Fse}), numerical calc.        & 0.001  & 0.012  & 0.012   \\
Sum                                               & -0.031 & -0.170 & -0.169  \\
                                                  &        &        &         \\
Eq.~(\ref{FQED}), numerical                       & -0.031 & -0.170 & -0.169  \\
\hline
\hline
\end{tabular}%
}
\end{table}

The values of the $F^{\rm SE}$ contribution to the FS constant using the three approaches described above are given in Table~\ref{TabFqedAl} for the different levels of electronic correlation effects treatment: DHF, CCSD, and CCSD(T). One can see good agreement between the results obtained within these three approaches at each level. Moreover, one can see that electron correlation effects are crucial for this treatment. As can be seen from Table~\ref{TabFqedAl}, the dominant SE contribution to the FS constant comes from the first term of Eq.~(\ref{Fse}), which implies a direct differentiation of the SE matrix elements. The second term, which does not include such a derivative, contributes more than an order of magnitude less.

As noted in section \ref{sec:recoilQED}, two types of QED contributions to the nuclear recoil effect were considered: (i) the effect beyond the approximation given by Eq.~(\ref{opNMS}), and (ii) the change of the NMS and SMS operators expectation values due to the perturbation of the electronic wave function by the SE and VP interactions. Table \ref{TabRecoilQED} provides the values of the QED contribution  of first type to the normal mass shift constant $k_{\text{NMS}}$ calculated at the different levels of theory: DHF, CCSD, and CCSD(T). As in the case of the SE contribution to the FS constant, one can see a significant role of the electron correlation effects. The related QED contribution to $k_{\text{SMS}}$ was estimated to be about an order of magnitude smaller using the rigorous QED treatment similar to Ref.~\cite{Anisimova:2022}. The QED contribution of the second type to the normal mass shift constant $k_{\text{NMS}}$ was found to be smaller by about a factor of 5.5 and of opposite sign compared to the first contribution being about $-0.2$~GHz~u for all the considered transitions. In addition, we found that the second-type QED contributions to $k_{\text{NMS}}$ and $k_{\text{SMS}}$ have similar absolute values but opposite signs, leading to a near cancellation of these effects in the total nuclear recoil factor $k_{\text{NMS}} + k_{\text{SMS}}$.

\begin{table}[h]
\caption{Values of the QED contribution to the nuclear recoil effect (in GHz~u) considered at different levels of theory.}
\label{TabRecoilQED}
\begin{tabular}{l
                S[table-format=3.3, table-align-text-post=false]
                S[table-format=2.2, table-align-text-post=false]
                S[table-format=2.2, table-align-text-post=false]}
\hline
\hline
Transtion                            & {DHF} & {CCSD} & {CCSD(T)} \\
\hline              
$3s^23p~^2P_{1/2} \to 3s^24s~^2S_{1/2}$ & 0.2 & 1.2  & 1.2     \\
$3s^23p~^2P_{3/2} \to 3s^24s~^2S_{1/2}$ & 0.2 & 1.2  & 1.2     \\
$3s^23p~^2P_{1/2} \to 3s^25s~^2S_{1/2}$ & 0.1 & 1.1  & 1.1     \\
$3s^23p~^2P_{3/2} \to 3s^25s~^2S_{1/2}$ & 0.1 & 1.1  & 1.1     \\
\hline
\hline
\end{tabular}
\end{table}

As one can see  from Table~\ref{Tenergies}, the basis set and the Breit interaction corrections also have opposite signs and similar absolute values for $k_{\text{NMS}}$ and $k_{\text{SMS}}$.
Thus, these corrections nearly cancel each other in the total nuclear recoil atomic factor.
However, to be conservative, we calculated the uncertainty of the total recoil effect as the square root of the sum of the squares of the uncertainties for $k_{\text{NMS}}$ and $k_{\text{SMS}}$, i.e., we did not employ this cancellation. Note also that we did not find such a cancellation for higher-order correlation effects, such as the contributions of quadruple excitations.

In Ref.~\cite{Plattner:2023} the IS shift $\Delta \nu^{27,26m}=377.5(3.4)$ MHz was measured for the $3s^23p~^2P_{3/2} \to 3s^24s~^2S_{1/2}$ transition. Using the previously calculated atomic factors $F=76.2(2.2)$~MHz/fm$^2$,  $k_{\rm MS}$=234(4) GHz~u ~\cite{Filippin:2016,Heylen:2021} and nucleus masses from Refs.~\cite{Huang:2021,AlkemadeAl26m:1982} the $\delta\langle r^2\rangle^{27,26m}$=0.429(45)(76) fm$^2$ value was deduced, where the first uncertainty is the experimental one and the second is due to the previously used atomic factors~\cite{Filippin:2016,Heylen:2021}. With the new atomic factors we obtain from Eq.~(\ref{freq})
\begin{equation}
\label{deducedr2}
\delta\langle r^2\rangle^{27,26m}=0.443(44)(19)~{\rm fm}^2.    
\end{equation}
The theoretical uncertainty of $\delta\langle r^2\rangle^{27,26m}$ is reduced by a factor of 4 and now the uncertainty of the deduced ms charge radius is dominated by the experimental uncertainty. 

To obtain the absolute ms charge radius of $^{26m}$Al, $R_c(^{26m}{\rm Al})$, the absolute ms charge radius of $^{27}$Al is required. The latter was tabulated in Ref.~\cite{Heylen:2021}, $R_c(^{27}{\rm Al})=$3.061(6)~fm, where the experimental data from the muonic atom spectroscopy were combined with elastic electron scattering measurements. Using this result together with the deduced value of $\delta\langle r^2\rangle^{27,26m}$ (\ref{deducedr2}) we obtain the absolute rms nucleus charge radius of $^{26m}$Al:
\begin{equation}
    R_c(^{26m}{\rm Al})=3.132(10)~{\rm fm}.
\end{equation}
This value is in good agreement with the previously derived value $R_c(^{26m}{\rm Al})=3.130(15)$~fm~\cite{Plattner:2023}, but 1.5 times more accurate due to the improved accuracy of the IS atomic factors. Now the uncertainty of the absolute rms charge radius is dominated by the uncertainties of the measured IS and $R_c(^{27}{\rm Al})$ and not by the uncertainty of the calculated atomic factors. The $\delta\langle r^2\rangle^{27,26m}$ and $R_c(^{26m}{\rm Al})$ are summarized in Table~\ref{Tradii}.

\begin{table}[]
\caption{Values of $\delta\langle r^2\rangle^{27,26m}$ (in ${\rm fm}^2$) and $R_c(^{26m}{\rm Al})$ (in ${\rm fm}$).
The first uncertainty for $\delta\langle r^2\rangle^{27,26m}$ is the experimental one and the second is due to the atomic factors.}
\begin{tabular}{lrr}
\hline
\hline
 & ~~~~$\delta\langle r^2\rangle^{27,26m}$ & ~~~~$R_c(^{26m}{\rm Al})$  \\
\hline 
Refs.~\cite{Towner:2002,Towner:2008} &    &   3.040(20) \\
Ref.~\cite{Plattner:2023} &  0.429(45)(76)  &  3.130(15)  \\
This work &  0.443(44)(19)  & 3.132(10)   \\
\hline
\hline
\end{tabular}
\label{Tradii}
\end{table}

By using the deduced $\delta\langle r^2\rangle^{27,26m}$ value together with the computed FS and MS atomic factors, one can determine the theoretical uncertainty for the IS of the transition of interest for the  $^{26m,27}$Al pair. The resulting uncertainty, 0.4\%, is dominated by the uncertainty of the mass shift constant. We can estimate the nonlinear (in $m/M$) mass shift effect contribution to the IS for the $^{26m,27}$Al pair by comparing the IS values calculated using Eq.~(\ref{freq}), i.e. the linear in $m/M$ effect, with the recoil effect calculated by including the operators (\ref{opNMS}) and (\ref{opSMS}) into the electronic Hamiltonian and performing direct calculations within the CCSD method for two masses: $M(^{26m}{\rm Al})$ and $M(^{27}{\rm Al})$. In the latter case we obtain an estimation of the nuclear recoil effect in all orders in $m/M$. According to this calculation, the nonlinear contribution of the  nuclear recoil effect was found to be 0.08\% of the total linear in $m/M$ mass shift value, i.e. about a factor of five smaller than the estimated nuclear-recoil-effect theoretical uncertainty for the considered transition (see Table \ref{AlISfactors}).

\begin{table*}[]
\caption{
Changes in ms charge radii of Al isotopes with respect to $^{27}$Al (neutron number $N$=14), ~$\delta\langle r^2\rangle^{27,A}$, and the absolute $\langle r^2\rangle$ values along the Al isotopic chain extracted from the IS measurements~\cite{Heylen:2021} for the $3s^23p~^2P_{3/2} \to 3s^24s~^2S_{1/2}$ transition. Uncertainties arising from the IS measurement~\cite{Heylen:2021}, beam energy~\cite{Heylen:2021} and atomic calculations are indicated for ~$\delta\langle r^2\rangle^{27,A}$ with rounds, square and curly parentheses, respectively. For $\langle r^2\rangle$ the total uncertainty is given.
}
\begin{tabular}{llcccc}
\hline
\hline
$A$~~~  & $N$~~~  & ~~~$\delta\langle r^2\rangle^{27,A}$, Ref.~\cite{Heylen:2021}~~~~~~~~~    & $\delta\langle r^2\rangle^{27,A}$, this work & ~~~$\langle r^2\rangle$, Ref.~\cite{Heylen:2021}~~~~~~~~~    & $\langle r^2\rangle$, this work                \\
\hline
28 & 15 & 0.003(10){[}43{]}\{72\}   & -0.013(10){[}43{]}\{18\} & 9.373(91)   & 9.357(60)     \\
29 & 16 & 0.142(8){[}84{]}\{134\}   & 0.110(8){[}83{]}\{35\}   & 9.511(163)  & 9.480(98)     \\
30 & 17 & 0.164(15){[}132{]}\{196\} & 0.119(16){[}132{]}\{51\} & 9.534(239)  & 9.489(146)    \\
31 & 18 & 0.301(16){[}178{]}\{250\} & 0.242(16){[}177{]}\{65\} & 9.671(311)  & 9.612(193)    \\
32 & 19 & 0.12(9){[}22{]}\{31\}     & 0.05(9){[}22{]}\{8\}     & 9.490(391)  & 9.422(255)  \\
\hline
\hline
\end{tabular}
\label{OtherRadii}
\end{table*}

In Ref.~\cite{Viatkina:2023} a detailed treatment of effects beyond the considered approximation (\ref{freq}), such as the higher order FS, the nuclear polarization and cross-terms of the FS and MS effects, was carried out for several isotopes of the slightly heavier element Ca ($Z=20$). In that work, it was found that the nuclear polarization effect contribution to the IS of the $4s \to 4p_{1/2}$ transition in Ca$^+$ is 
smaller than the nonlinear mass shift contribution.
It reached 0.03\% of the total IS value. Contributions of other effects were even smaller. This is more than an order of magnitude smaller than the present uncertainty. Thus, such effects were beyond the scope of present study.

\begin{figure}[!htb]
\centering
\includegraphics[width=0.5\textwidth]{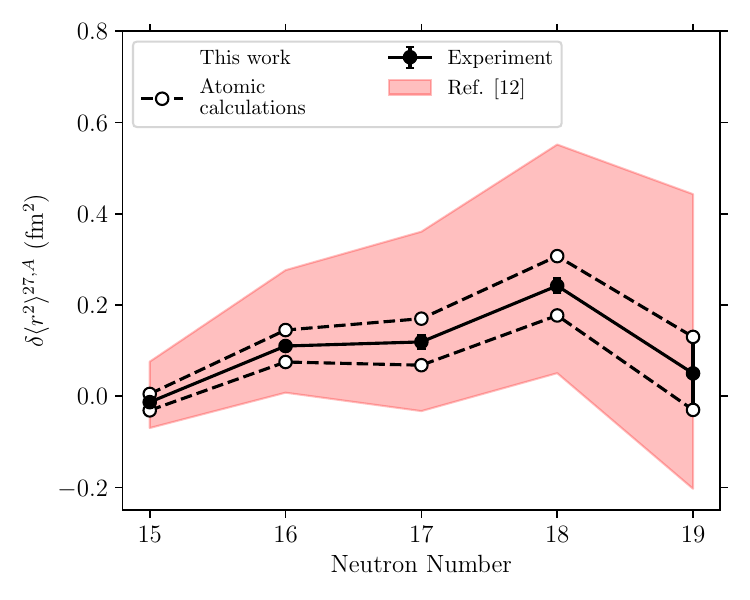}
\caption{\label{fig:dr} Changes in the mean-square charge radii of Al isotopes. The uncertainties arising from the atomic factors used in Ref. \cite{Heylen:2021} are represented by the shaded red area. Uncertainties associated to the beam  energy are not included. }
\end{figure}

As a final application of the calculated atomic factors for the $3s^23p~^2P_{3/2} \to 3s^24s~^2S_{1/2}$ transition, we derived  relative, $\delta\langle r^2\rangle^{27,A}$, and absolute ms charge radii for Al isotopes with mass numbers $A$=28--32 using the measured IS values from Ref.~\cite{Heylen:2021} and the absolute ms charge radius of $^{27}$Al~\cite{ANGELI201369,Heylen:2021}. The results for the $\delta\langle r^2\rangle^{27,A}$ are shown in Table~\ref{OtherRadii}, and compared with literature values in Fig. \ref{fig:dr}.

As shown in Fig. \ref{fig:dr}, the uncertainty resulting from atomic calculations of $\delta\langle r^2\rangle^{27,A}$  has now been reduced by a factor of 4 for, and the total uncertainty is now dominated by the uncertainty in the beam energy~\cite{Heylen:2021}, rather than by atomic calculations.
The total uncertainties of the ms charge radii of all considered Al isotopes are reduced by a factor of about 1.5. 

In the present paper we calculated the IS atomic factors for three other transitions  with similar accuracy, see Table~\ref{AlISfactors}. These data can be used for the interpretation of future experiments. The $3s^23p~^2P_{1/2} \to 3s^25s~^2S_{1/2}$  and $3s^23p~^2P_{3/2} \to 3s^25s~^2S_{1/2}$ transitions are of particular interest because the NMS and SMS contributions almost exactly cancel each other resulting in a significantly reduced nuclear recoil effect contribution to the IS.

\section{Conclusion}

We have developed the effective method to incorporate the QED contributions into the field shift atomic constant in neutral many-electron systems. The model operator used to calculate the QED effects on the nuclear recoil IS factor, as described in~\cite{Anisimova:2022}, has been adapted for the atomic IS correlation calculations. These QED terms have proven to be significant at our current level of precision achieved for Al, surpassing the uncertainties inherent in electronic calculations. These advancements can now be applied in other studies involving calculations of the IS atomic factors. Importantly, the developed procedure can also be applied for calculating FS factors in molecules.

As a result of these developments, we were able to reduce the uncertainty of the field shift atomic parameter in Al by more than an order of magnitude and the uncertainty of the nuclear recoil parameter by a factor of 4 compared to the previous studies~\cite{Filippin:2016,Heylen:2021}. By combining the measured IS with the obtained values of the atomic factors, the final theoretical uncertainty of the charge radius of $^{26m}$Al has been reduced by a factor of 1.5 compared to the previous study~\cite{Plattner:2023}.
Similar results were also derived for the $^{28}$Al, $^{29}$Al, $^{30}$Al, $^{31}$Al and $^{32}$Al isotopes using existing IS measurements~\cite{Heylen:2021}.

\begin{acknowledgments}  
The authors are grateful to V.M. Shabaev and D.E. Maison for valuable  discussions.

Electronic structure calculations have been carried out using computing resources of the federal collective usage center Complex for Simulation and Data Processing for Mega-science Facilities at National Research Centre ``Kurchatov Institute'', http://ckp.nrcki.ru/.

Electronic structure calculations were supported by the Russian Science Foundation (Grant No. 19-72-10019-P (https://rscf.ru/en/project/22-72-41010/).
Calculations of the self energy matrix elements within the rigorous QED approach were supported by the Foundation for the Advancement of Theoretical Physics and Mathematics BASIS (Project No. 21-1-3-52-1). R.F.G.R,  and F.P. acknowledge support from
the U.S. Department of Energy under the grants DE-SC0021176. A.B. acknowledge support from
the National Science Foundation of the US, FAIN 2207996.

\end{acknowledgments}


%

\end{document}